\documentclass[aps,prd,twocolumn,superscriptaddress,amsmath,showpacs,floatfix,nofootinbib]{revtex4-1}
\usepackage{graphicx}
\usepackage{dcolumn}
\usepackage{bm}
\usepackage{natbib}
\usepackage{multirow}

\newcommand{\Mpc}{{\rm ~Mpc}}

\begin{document}

\title{Neutrino Anisotropies after Planck}

\author{Martina Gerbino} \email{martina.gerbino@roma1.infn.it}
\affiliation{Physics Department and INFN, Universit\`a di Roma ``La Sapienza'', Ple Aldo Moro 2, 00185, Rome, Italy}

\author{Eleonora Di Valentino} \email{eleonora.divalentino@roma1.infn.it}
\affiliation{Physics Department and INFN, Universit\`a di Roma ``La Sapienza'', Ple Aldo Moro 2, 00185, Rome, Italy}

\author{Najla Said} \email{saidnajl@roma1.infn.it}
\affiliation{Physics Department and INFN, Universit\`a di Roma ``La Sapienza'', Ple Aldo Moro 2, 00185, Rome, Italy}

\begin{abstract}
We present new constraints on the rest-frame sound speed, $c_{\rm eff}^2$, and the viscosity parameter, $c_{\rm vis}^2$,
of the Cosmic Neutrino Background from the recent measurements of the Cosmic Microwave Background anisotropies 
provided by the $Planck$ satellite. While broadly consistent with the expectations of $c_{\rm eff}^2=c_{\rm vis}^2=1/3$ in the standard scenario, the $Planck$ dataset hints at a higher
value of the viscosity parameter, with  $c_{\rm vis}^2=0.60\pm0.18$ at $68 \%$ c.l., and a lower value of the
sound speed, with $c_{\rm eff}^2=0.304\pm0.013$ at $68 \%$ c.l..
We find a correlation between the neutrino parameters and the lensing amplitude of the 
temperature power spectrum $A_{\rm L}$.  When the latter parameter is allowed to vary, we find a better
consistency with the standard model with $c_{\rm vis}^2=0.51\pm0.22$, 
$c_{\rm eff}^2=0.311\pm0.019$ and $A_{\rm L}=1.08\pm0.18$ at $68 \%$ c.l..
This result indicates that the anomalous large value of $A_{\rm L}$ measured by $Planck$ could be
connected to non-standard neutrino properties. Including additional datasets
from Baryon Acoustic Oscillation surveys and the Hubble Space Telescope constraint on the
Hubble constant, we obtain  $c_{\rm vis}^2=0.40\pm0.19$, 
$c_{\rm eff}^2=0.319\pm0.019$, and $A_{\rm L}=1.15\pm0.17$ at $68 \%$ c.l.; including the lensing power spectrum, we obtain $c_{\rm vis}^2=0.50\pm0.19$, $c_{\rm eff}^2=0.314\pm0.015$, and $A_{\rm L}=1.025\pm0.076$ at $68 \%$ c.l..
Finally, we investigate further degeneracies between the clustering parameters and other cosmological parameters.
\end{abstract}
 

\maketitle

\section{Introduction} \label {sec:intro}

The recent measurements of the Cosmic Microwave Background  (CMB, hereafter) anisotropies provided by the Planck satellite \cite{Ade:2013xsa, Planck:2013kta, Ade:2013lta} are in excellent agreement with expectations of the standard $\Lambda$CDM cosmological model 
and present the tightest ever constraints on its parameters.
These new observations open the opportunity to further test some of the assumptions of the $\Lambda$CDM model
and to possibly identify the presence of new physics.

Following this line of investigation and previous analyses (see e.g. \cite{najla,Archidiacono:2013lva,Archidiacono:2011gq,tristan,Hou:2011ec}),
 in this paper we test some properties of the Cosmic Neutrino
Background (CNB, hereafter). It is well known that, apart from CMB photons, the standard
cosmological model predicts the existence of a CNB
of energy density (when neutrinos are relativistic) of

\begin{equation}
\rho_{\rm rad} = \left[1+ \frac{7}{8} \left(\frac{4}{11}\right)^\frac{4}{3} N_{\rm eff} \right] \rho_{\gamma}\,
\end{equation}

where $\rho_{\gamma}$ is the photon energy density for $T_{\gamma}=2.725 \: K$ and $N_{\rm eff}$ is the number of relativistic degrees of freedom. 

In the past years, cosmological data increasingly constrained the value of $N_{\rm eff}$, ruling out the possibility of $N_{\rm eff}=0$ at high significance (e.g. see \cite{Archidiacono:2013lva} and references therein). More recently, the $Planck$ experiment \cite{Ade:2013lta} reported
the bound $N_{\rm eff}=3.51\pm0.39$ at $68\%$ c.l., consistent in between two standard deviations 
with the standard value $N_{\rm eff}=3.046$ and providing evidence for the neutrino
background at the level of about nine standard deviations.

However, the slightly higher value for $N_{\rm eff}$ suggests that new physics can be present in the neutrino sector.
This new physics could also be connected with the lensing amplitude of the temperature power spectrum.
Let us remind here that gravitational lensing acts on the CMB by deflecting the photon path by a quantity defined by the gradient of the lensing potential $\phi\left({\bf n}\right)$, integrated along the line of sight ${\bf n}$. Lensing also affects the power spectrum by smoothing the acoustic peaks and this effect can be parameterized by introducing the lensing amplitude parameter $A_{\rm L}$, as defined in \cite{Calabrese:2008rt}, which performs a rescaling of the lensing potential
\begin{equation}
C_{\ell}^{\phi \phi} \rightarrow A_{\rm L} C_{\ell}^{\phi \phi}
\end{equation}
where $C_{\ell}^{\phi \phi}$ is the power spectrum of the lensing field. 
Interestingly, the $Planck$ data suggest an anomalous value for the lensing amplitude 
of $A_{\rm L}=1.22^{+0.11}_{-0.13}$ at $68\%$ c.l., i.e. higher respect to the expected value of $A_{\rm L}=1$ at about two
standard deviations. 

Allowing a simultaneous variation in $N_{\rm eff}$ and $A_{\rm L}$, we recently found
$N_{\rm eff}=3.71\pm0.40$ and $A_{\rm L}=1.25\pm0.13$ at $68\%$ c.l. in \cite{najla}, 
suggesting the presence of some anomalies at higher significance.

In this paper, we continue our search for anomalies but considering a different modification to the CNB. 
Indeed, instead of varying the neutrino effective number, which we fix at $N_{\rm eff}=3.046$, 
we modify the CNB clustering properties as first proposed in \cite{Hu:1998tk}. 
Following \cite{Hu:1998kj}, the CNB can be modelled as a Generalized Dark Matter (GDM) component with a set of equations, describing the evolution of perturbations, given by (\cite{Archidiacono:2011gq, Hu:1998kj, Trotta:2004ty}):

\begin{align}\label{perturbations}
&\dot \delta_{\nu} =\frac{\dot a}{a} \left(1-3 c_{\rm eff}^2\right) \left(\delta_{\nu}+3\frac{\dot a}{a} \frac{q_{\nu}}{k}\right)-k\left(q_{\nu}+\frac{2}{3k}\dot h\right)\,\\
&\dot q_{\nu} = k \, c_{\rm eff}^2 \left(\delta_{\nu}+3\frac{\dot a}{a} \frac{q_{\nu}}{k}\right)-\frac{\dot a}{a} q_{\nu}-\frac{2}{3}k \pi_{\nu}\,\\
&\dot \pi_{\nu} = 3 \, c_{\rm vis}^2 \left(\frac{2}{5} q_{\nu} + \frac{8}{15}\sigma\right) - \frac{3}{5} k F_{\nu,3}\,\\
&\frac{2 l +1}{k} \dot F_{\nu,l} - l F_{\nu, l-1} = -\left(l+1\right) F_{\nu, l+1} \, \: l\ge 3\,
\end{align}

where $c_{\rm eff}^2$ is the sound speed in the CNB rest frame, describing pressure fluctuations respect to density perturbations, and $c_{\rm vis}^2$ is the ``viscosity'' parameter which parameterizes the anisotropic stress. For standard neutrinos, we have $c_{\rm eff}^2=c_{\rm vis}^2=1/3$. Constraints on these parameters have been set by several authors (see e.g. \cite{Archidiacono:2013lva, Archidiacono:2011gq, Diamanti:2012tg}), since the observation of deviations from the standard values could hint at non-standard physics. 
Using cosmological data previous to $Planck$, Smith $et$ $al.$ \cite{tristan} found that, assuming $N_{\rm eff}=3.046$, the case $c_{\rm eff}^2=c_{\rm vis}^2=1/3$ is ruled out at the level of two standard deviations. More recently, Archidiacono $et$ $al.$ \cite{Archidiacono:2011gq} found that current cosmological data from the South Pole Telescope SPT \cite{Hou:2012xq, Story:2012wx} exclude the standard value of $c_{\rm vis}^2=1/3$ at $2\sigma$ level, pointing towards a lower value. 

It is therefore extremely timely to bound the values of the neutrino perturbation parameters using the recent $Planck$ data.
Here we provide those new constraints but also considering the possible degeneracies between $c_{\rm eff}^2$, $c_{\rm vis}^2$
and the temperature power spectrum lensing amplitude $A_{\rm L}$.

The paper is organized as follows: the next section is devoted to the description of the analysis method; our results are summarized in Section III; conclusions are drawn in Section IV.

\begin{table*}[ht!]
\caption{Comparison between extended cosmological models and the standard $\Lambda$CDM for the $PlanckWP$ dataset. Listed are posterior means for the cosmological parameters from the indicated datasets (errors refer to 68\% credible intervals).}
\begin{ruledtabular}
\label{tab:paramsWP}
\footnotetext [1] {km s$^{-1}$ Mpc$^{-1}$}
\begin{tabular}{c|c|ccccc}
 Dataset &Parameter& $\Lambda$CDM & $+c_{\rm vis}^2+c_{\rm eff}^2$ & $+c_{\rm eff}^2+A_{\rm L}$ &$+c_{\rm vis}^2+A_{\rm L}$ & $+c_{\rm eff}^2+c_{\rm vis}^2+A_{\rm L}$ \\ \hline
 \multirow{12}{*}{{\bf PlanckWP}} &$100\,\Omega_b h^2$	
 &$2.206\pm0.028$ &$2.118\pm0.047$&$2.219\pm0.045$ &$2.236\pm0.053$ & $2.162\pm0.095$   \\
 &$\Omega_c h^2$        
 &$0.1199\pm0.0027$&$0.1157\pm0.0038$ &$0.1177\pm0.0032$ &$0.1170\pm0.0034$ & $0.1159\pm0.0036$ \\
 &$100\, \theta$          
 &$1.0413\pm0.0006$ &$1.0412\pm0.0014$ &$1.0428\pm0.0012$ &$1.0421\pm0.0019$ & $1.0420\pm0.0020$\\       
 &$\log[10^{10}A_S]$ \phantom{X}
 &$ 3.089\pm0.025$ &$3.173\pm0.052$ &$3.086\pm0.028$ &$3.08\pm0.05$ & $3.141\pm0.078$\\
 &$\tau$
 &$0.090\pm0.013$&$0.089\pm0.013$ &$0.088\pm0.013$  &$0.087\pm0.013$ & $0.089\pm0.014$\\
 &$n_S$
 &$0.9606 \pm 0.0073$&$0.998\pm0.018$ &$0.9732\pm0.0099$  &$0.970\pm0.014$ & $0.989\pm0.023$\\
&$A_{\rm L}$			
 &$\equiv1$&$\equiv1$ &$1.16\pm0.13$ &$1.20 \pm 0.12$ & $1.08\pm0.18$\\
 &$c_{\rm vis}^2$
 &$\equiv0.33$&$0.60\pm0.18$ &$\equiv0.33$ &$0.35 \pm 0.12$ & $0.51\pm0.22$\\
 &$c_{\rm eff}^2$			
 &$\equiv0.33$&$0.304 \pm 0.013$ &$0.321\pm0.014$ &$\equiv0.33$ & $0.311\pm0.019$\\
 &$H_0$ \ \ $^{(a)}$
 &$67.3\pm1.2$ &$68.0\pm1.3$ &$68.7\pm1.5$	&$68.9\pm1.5$	& $68.6\pm1.7$\\
\end{tabular}
\end{ruledtabular}
\end{table*}

\section{Data Analysis Method} \label{sec:method}

We sample a four-dimensional set of standard cosmological parameters, imposing flat priors: the baryon and cold dark matter densities $\Omega_ b$ and $\Omega_c$, the angular size of the sound horizon at decoupling $\theta$, and the optical depth to reionization $\tau$. As inflationary parameters we consider the scalar spectral index $n_S$ and the overall normalization of the spectrum $A_S$ at $k=0.05\Mpc^{-1}$.
We consider purely adiabatic initial conditions and we impose spatial flatness. As far as neutrino components are concerned, we assume the $Planck$ collaboration \cite{Ade:2013lta} baseline model: the total number of relativistic degrees of freedom $N_{\rm eff}=3.046$, with a single massive eigenstate with $m_{\nu}=0.06 \: \rm eV$. We checked that assuming massless neutrinos does not affect our results.
The Helium abundance $Y_p$ is also varied but assuming a Big Bang Nucleosynthesis (BBN) consistency (given $N_{\rm eff}$ and $\Omega_b$, $Y_p$ is a determined function of them) and is therefore not treated as an extra free parameter.

When exploring extended cosmological scenarios, we vary the effective sound speed $c_{\rm eff}^2$, the viscosity parameter $c_{\rm vis}^2$ and the lensing amplitude parameter $A_{\rm L}$. Firstly, we combine them in pair ($c_{\rm eff}^2-c_{\rm vis}^2$, $c_{\rm eff}^2-A_{\rm L}$, $c_{\rm vis}^2-A_{\rm L}$), fixing the third parameter at its standard value ($A_{\rm L}=1$, $c_{\rm vis}^2=1/3$, $c_{\rm eff}^2=1/3$). Finally, we combine the three parameters all together. We adopt the following flat priors: $0 \le c_{\rm vis}^2, c_{\rm eff}^2 \le 1$ and $0 \le A_{\rm L} \le 4$.

Concerning the datasets, we use the high-$\ell$ $Planck$ temperature power spectrum (\texttt{CamSpec}, $50 \le \ell \le 2500$ \cite{Planck:2013kta}) and the low-$\ell$ $Planck$ temperature power spectrum (\texttt{commander}, $2 \le \ell \le 49$ \cite{Planck:2013kta}), in combination with the $WMAP$ low-$\ell$ likelihood for polarization \cite{Bennett:2012fp} at $\ell \leq 23$. We refer to this combination as $PlanckWP$.

We also consider the inclusion of additional datasets.  Referring to the latest measurements of the Hubble Space Telescope (HST) \cite{hst}, we assume a gaussian prior on the Hubble constant $H_0=73.8\pm2.4 \,\mathrm{km}\,\mathrm{s}^{-1}\,\mathrm{Mpc}^{-1}$. We also include measurements of Baryon Acoustic Oscillations (BAO) from galaxy surveys. Here, we follow the same approach reported in Ref.\ \cite{Ade:2013lta} combining three datasets: SDSS-DR7 \cite{padmanabhan/etal:2012}, SDSS-DR9 \cite{Anderson:2013oza} and $\rm 6dF$ Galaxy Survey \cite{Beutler:2011hx}. We refer to this combination as $PlanckEX$.

Since the information on the lensing amplitude derived from the trispectrum do not hint at the high value of the lensing parameter allowed by the temperature power spectrum (see discussion in \cite{Ade:2013lta}), we also investigate the impact of the addition of the $Planck$ \texttt{lensing\_likelihood} \cite{Ade:2013tyw} to the $PlanckWP$ dataset. We refer to this combination as $PlanckL$.

Our analysis method is based on the publicly available Monte Carlo Markov Chain package \texttt{cosmomc} \cite{Lewis:2002ah, Lewis:2013hha} (version released in March 2013) using the Gelman and Rubin statistic as convergence diagnostic.

\begin{table*}[htb!]
\caption{Comparison between extended cosmological models and the standard $\Lambda$CDM for the $PlanckEX$ dataset. Listed are posterior means for the cosmological parameters from the indicated datasets (errors refer to 68\% credible intervals).}
\begin{ruledtabular}
\label{tab:paramsEX}
\footnotetext [1] {km s$^{-1}$ Mpc$^{-1}$}
\begin{tabular}{c|c|ccccc}
 Dataset &Parameter& $\Lambda$CDM & $+c_{\rm vis}^2+c_{\rm eff}^2$ & $+c_{\rm eff}^2+A_{\rm L}$ &$+c_{\rm vis}^2+A_{\rm L}$ & $+c_{\rm eff}^2+c_{\rm vis}^2+A_{\rm L}$ \\ \hline
\multirow{12}{*}{{\bf PlanckEX}} &$100\,\Omega_b h^2$	
 &$2.236\pm0.026$ &$2.142\pm0.048$ &$2.228\pm0.037$ &$2.258\pm0.043$ & $2.209\pm0.091$  \\
 &$\Omega_c h^2$        
 &$0.1180\pm0.0016$&$0.1160\pm0.0025$ &$0.1171\pm0.0018$ &$0.1173\pm0.0023$ & $0.1167\pm0.0024$ \\
 &$100\, \theta$          
 &$1.0417\pm0.0006$ &$1.0417\pm0.0016$ &$1.0427\pm0.0011$ &$1.0429\pm0.0018$ & $1.0426\pm0.0018$\\       
 &$\log[10^{10}A_S]$ \phantom{X}
 &$ 3.111\pm0.032$ &$3.158\pm0.050$ &$3.084\pm0.028$ &$3.065\pm0.043$ & $3.104\pm0.075$\\
 &$\tau$
 &$0.101\pm0.015$&$0.091\pm0.013$ &$0.088\pm0.013$  &$0.087\pm0.013$ & $0.088\pm0.013$\\
 &$n_S$
 &$0.9621 \pm 0.0061$&$0.993\pm0.016$  &$0.9737\pm0.0080$ &$0.967\pm0.012$ & $0.978\pm0.023$\\
&$A_{\rm L}$			
 &$\equiv1$&$\equiv1$ &$1.18\pm0.12$ &$1.23\pm0.11$ & $1.15\pm0.17$\\
 &$c_{\rm vis}^2$
 &$\equiv0.33$ &$0.53\pm0.16$ &$\equiv0.33$ &$0.302\pm0.097$ & $0.40\pm0.19$ \\
 &$c_{\rm eff}^2$			
 &$\equiv0.33$&$0.306\pm 0.013$ &$0.322\pm0.013$ &$\equiv0.33$ & $0.319\pm0.019$\\
 &$H_0$ \ \ $^{(a)}$
 &$68.31\pm0.73$ &$68.29\pm0.74$ &$68.93\pm0.80$ &$69.16\pm0.87$	& $68.88\pm0.99$\\
\end{tabular}
\end{ruledtabular}
\end{table*}

\begin{table*}[htb!]
\caption{Comparison between extended cosmological models and the standard $\Lambda$CDM for the $PlanckL$ dataset. Listed are posterior means for the cosmological parameters from the indicated datasets (errors refer to 68\% credible intervals).}
\begin{ruledtabular}
\label{tab:paramsL}
\footnotetext [1] {km s$^{-1}$ Mpc$^{-1}$}
\begin{tabular}{c|c|ccccc}
 Dataset &Parameter& $\Lambda$CDM & $+c_{\rm vis}^2+c_{\rm eff}^2$ & $+c_{\rm eff}^2+A_{\rm L}$ &$+c_{\rm vis}^2+A_{\rm L}$ & $+c_{\rm eff}^2+c_{\rm vis}^2+A_{\rm L}$ \\ \hline
 \multirow{12}{*}{{\bf PlanckL}} &$100\,\Omega_b h^2$	
 &$2.206\pm0.028$ &$2.155\pm0.049$&$2.214\pm0.038$ &$2.217\pm0.046$ & $2.166\pm0.063$   \\
 &$\Omega_c h^2$        
 &$0.1199\pm0.0027$&$0.1151\pm0.0034$ &$0.1171\pm0.0030$ &$0.1159\pm0.0034$ & $0.1151\pm0.0037$ \\
 &$100\, \theta$          
 &$1.0413\pm0.0006$ &$1.0414\pm0.0015$ &$1.0426\pm0.0012$ &$1.0415\pm0.0018$ & $1.0415\pm0.0017$\\       
 &$\log[10^{10}A_S]$ \phantom{X}
 &$ 3.089\pm0.025$ &$3.146\pm0.054$ &$3.084\pm0.027$ &$3.094\pm0.051$ & $3.137\pm0.064$\\
 &$\tau$
 &$0.090\pm0.013$&$0.090\pm0.013$ &$0.088\pm0.013$  &$0.088\pm0.013$ & $0.088\pm0.013$\\
 &$n_S$
 &$0.9606 \pm 0.0073$&$0.990\pm0.019$ &$0.9726\pm0.0098$  &$0.974\pm0.015$ & $0.989\pm0.021$\\
&$A_{\rm L}$			
 &$\equiv1$&$\equiv1$ &$1.042\pm0.072$ &$1.057\pm0.070$ & $1.025\pm0.076$\\
 &$c_{\rm vis}^2$
 &$\equiv0.33$&$0.52\pm0.18$ &$\equiv0.33$ &$0.39\pm0.14$ & $0.50\pm0.19$\\
 &$c_{\rm eff}^2$			
 &$\equiv0.33$&$0.312\pm0.013$ &$0.322\pm0.012$ &$\equiv0.33$ & $0.314\pm0.015$\\
 &$H_0$ \ \ $^{(a)}$
 &$67.3\pm1.2$ &$68.5\pm1.1$ &$68.8\pm1.4$	&$68.9\pm1.4$	& $68.8\pm1.5$\\
\end{tabular}
\end{ruledtabular}
\end{table*}

\begin{figure*}[htb!]
\includegraphics[width=0.44\linewidth,keepaspectratio]{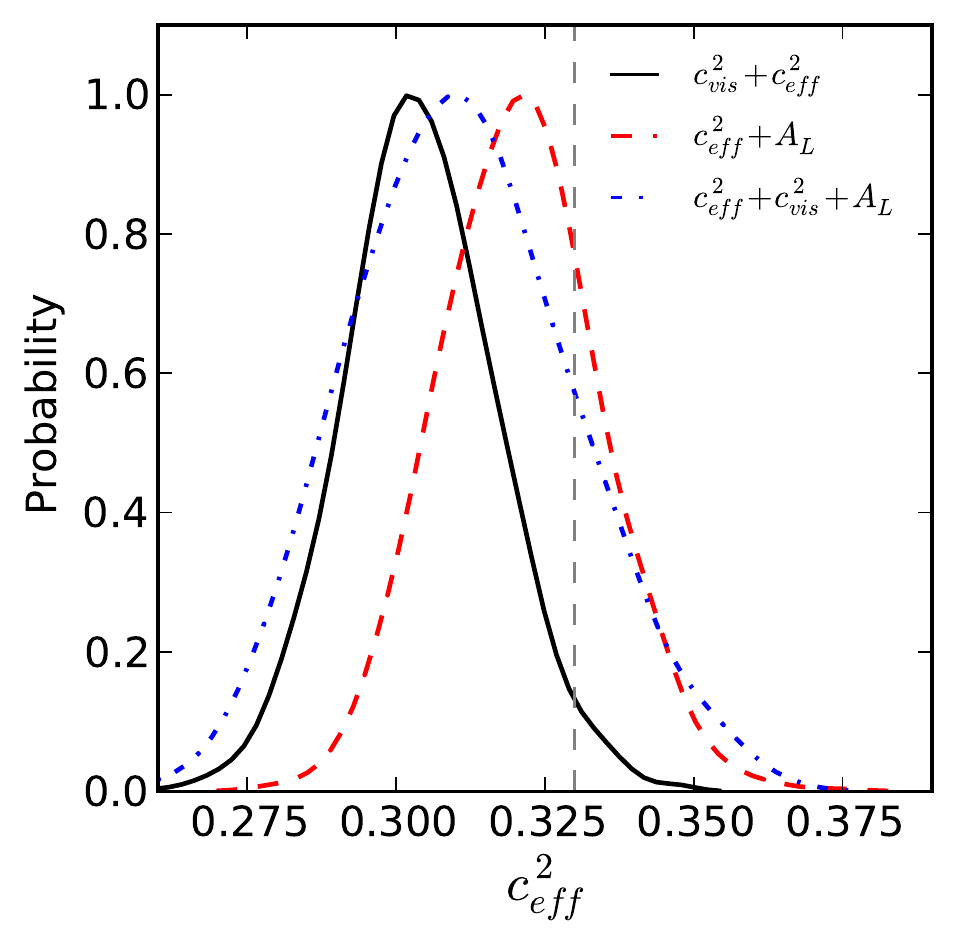}
\includegraphics[width=0.44\linewidth,keepaspectratio]{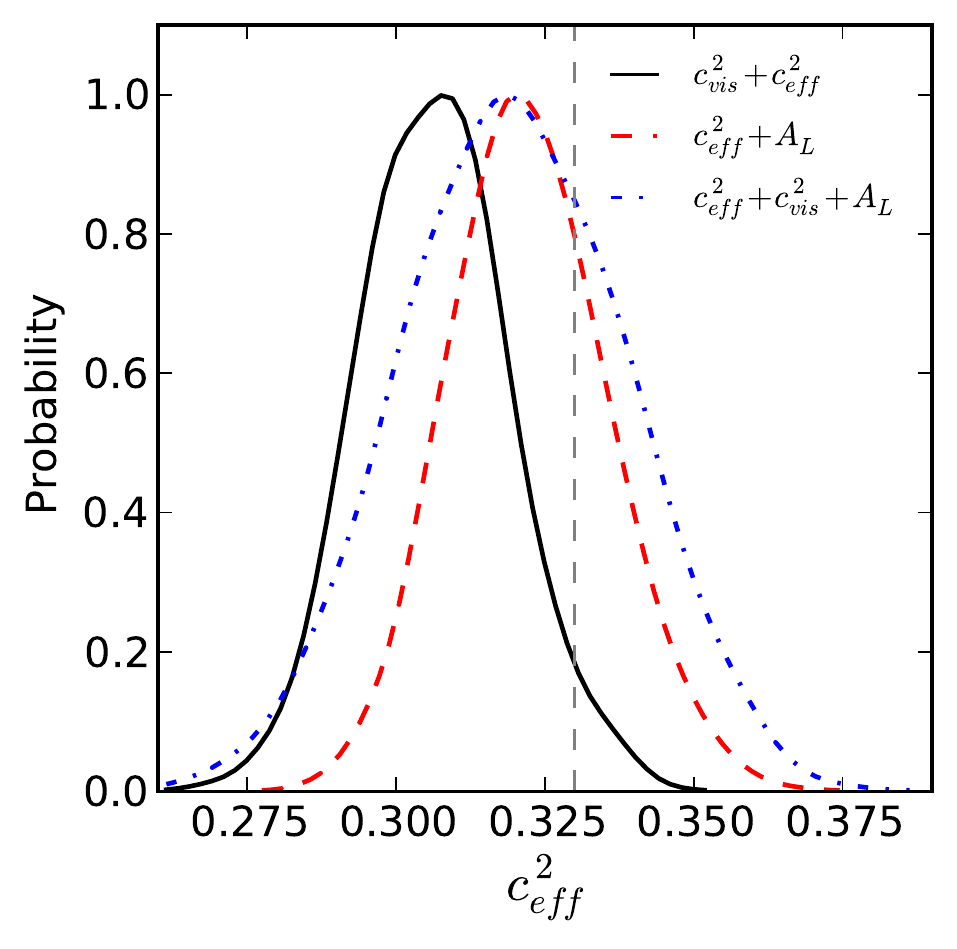}
\includegraphics[width=0.44\linewidth,keepaspectratio]{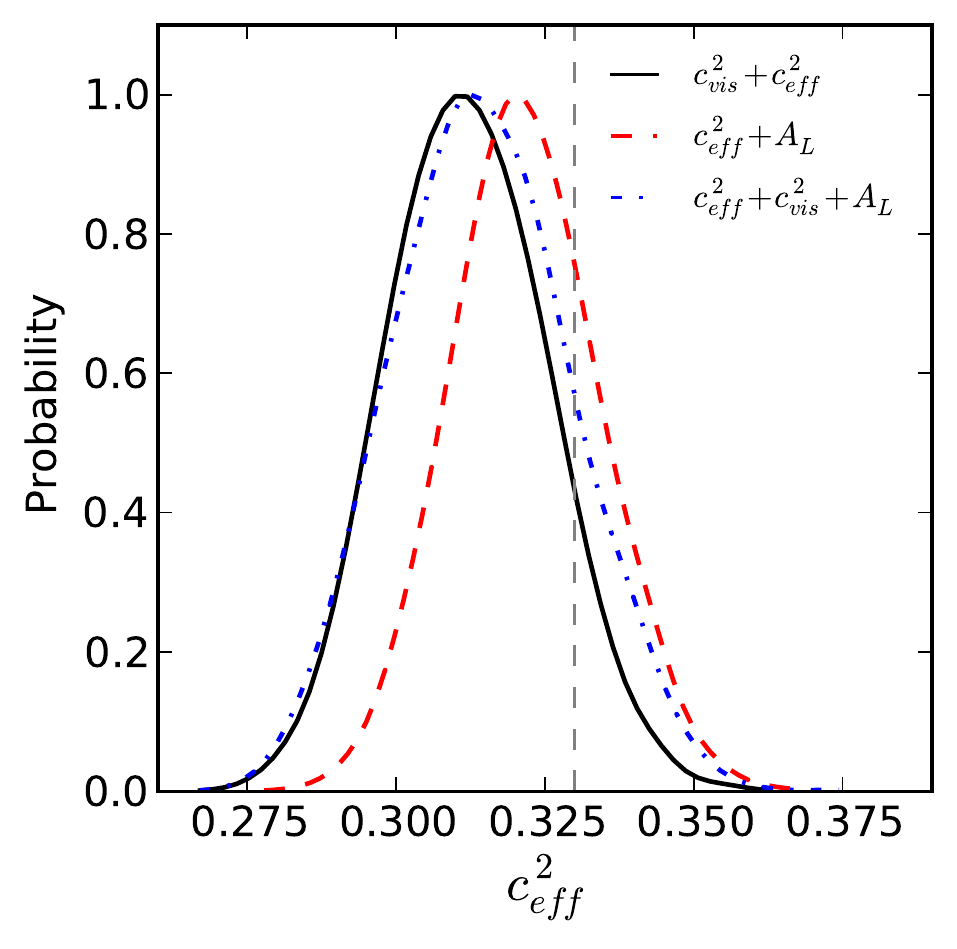}\\ 
\caption{One-dimensional posterior probabilities of the parameter $c_{\rm eff}^2$ for the indicated models for $PlanckWP$ (upper left), $PlanckEX$ (upper right) and $PlanckL$ (bottom) datasets. The vertical dashed line indicates the expected value in the standard model. Note the different range in the $x$ axes for the $PlanckL$ dataset.}

\label{fig:ceff_1D}
\end{figure*}

\begin{figure*}[htb!]
\includegraphics[width=0.44\linewidth,keepaspectratio]{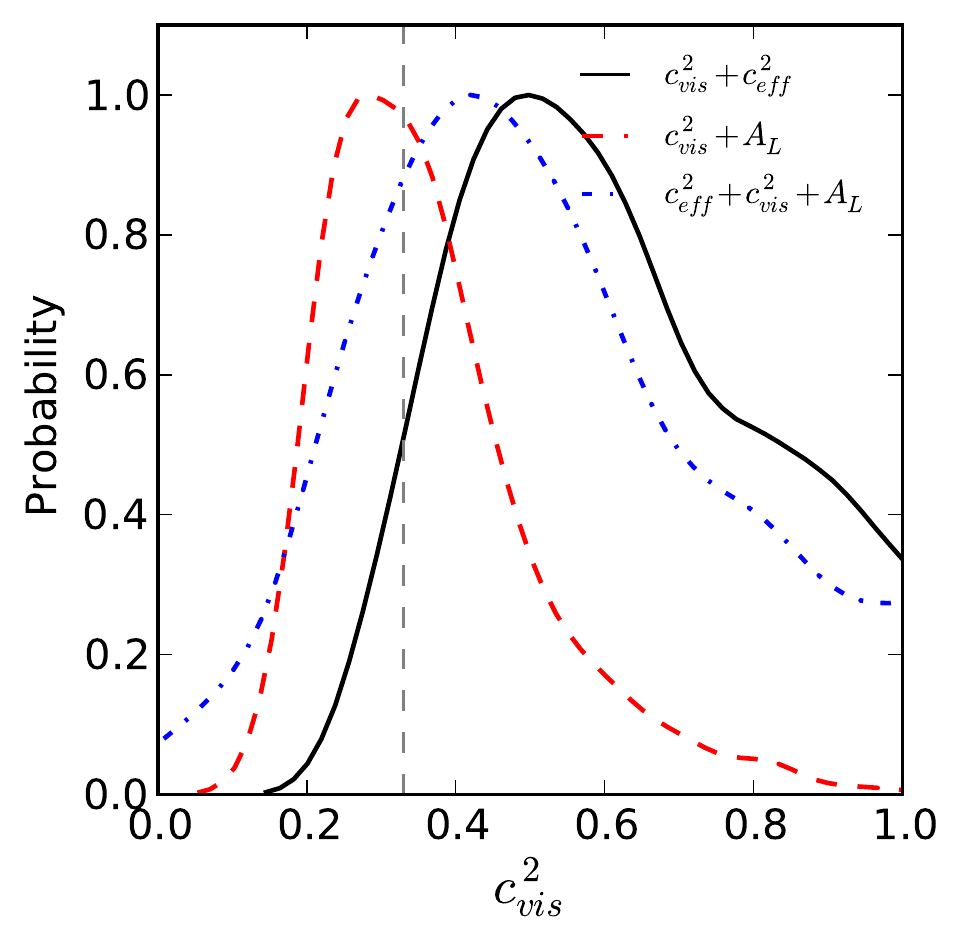}
\includegraphics[width=0.44\linewidth,keepaspectratio]{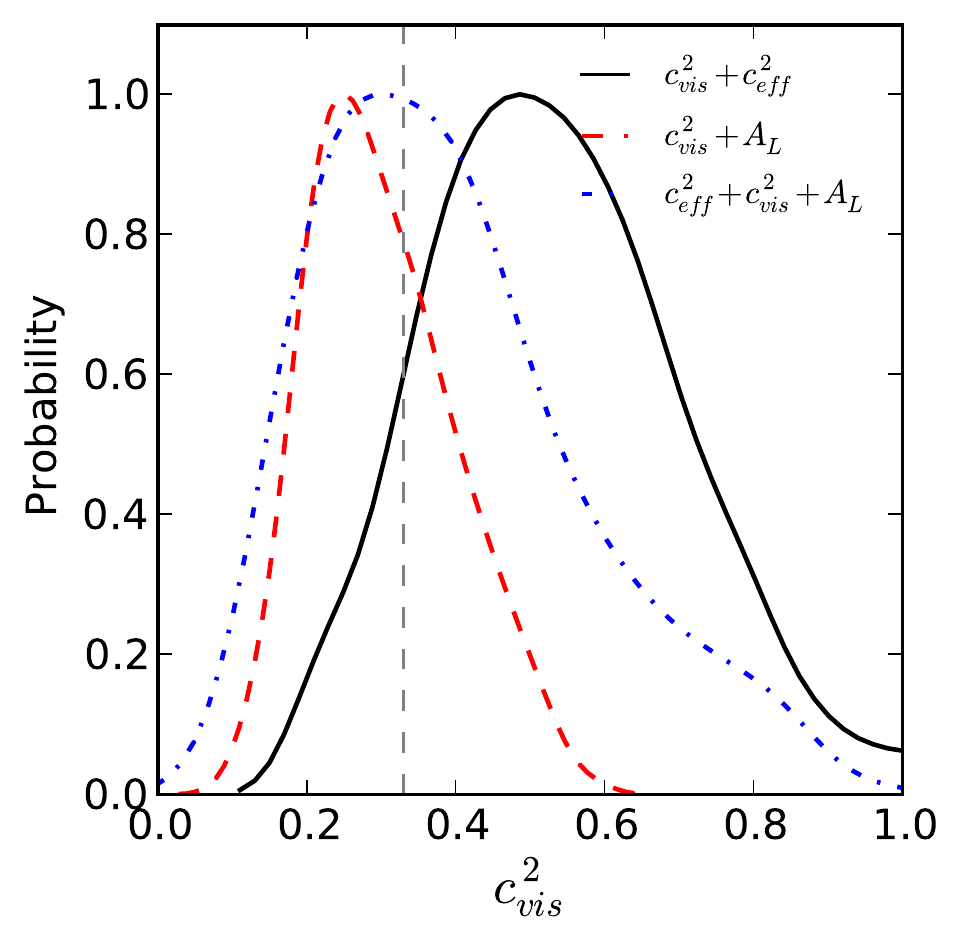}
\includegraphics[width=0.44\linewidth,keepaspectratio]{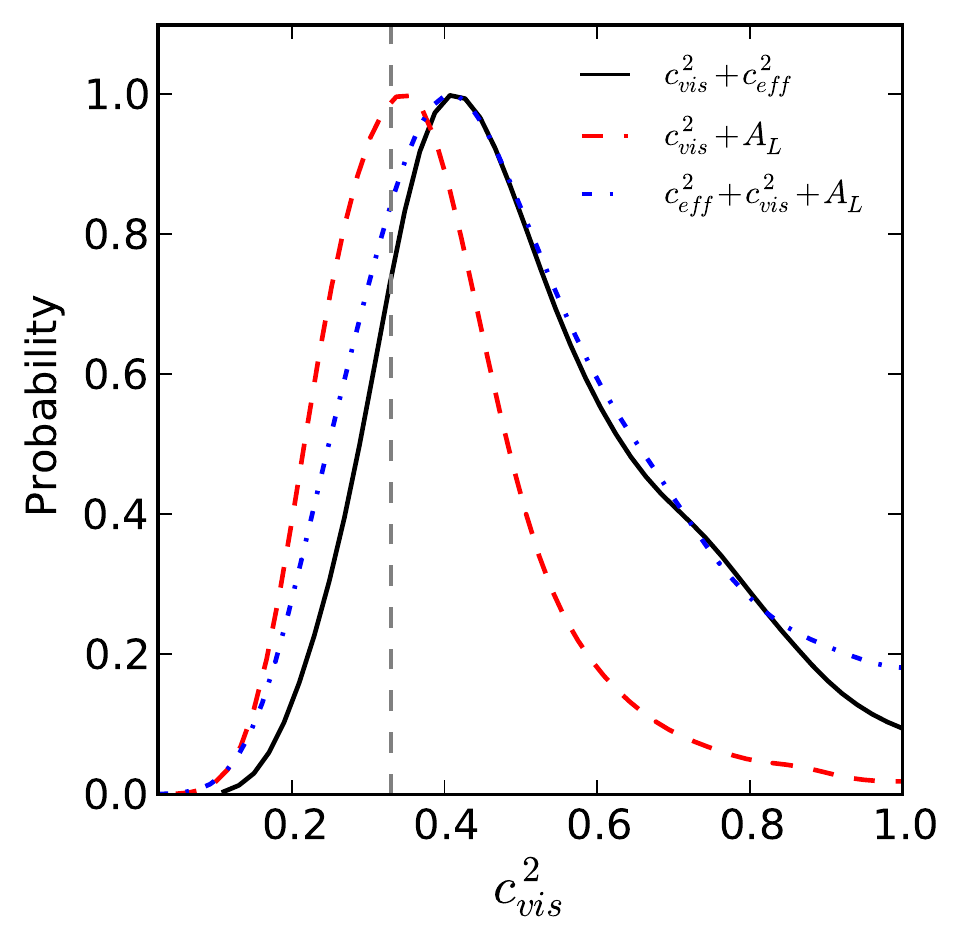} \\
\caption{One-dimensional posterior probabilities of the parameter $c_{\rm vis}^2$ for the indicated models for $PlanckWP$ (upper left), $PlanckEX$ (upper right) and $PlanckL$ (bottom) datasets. The vertical dashed line indicates the expected value in the standard model.}

\label{fig:cvis_1D}
\end{figure*}

\begin{figure*}[htb!]
\includegraphics[width=0.44\linewidth,keepaspectratio]{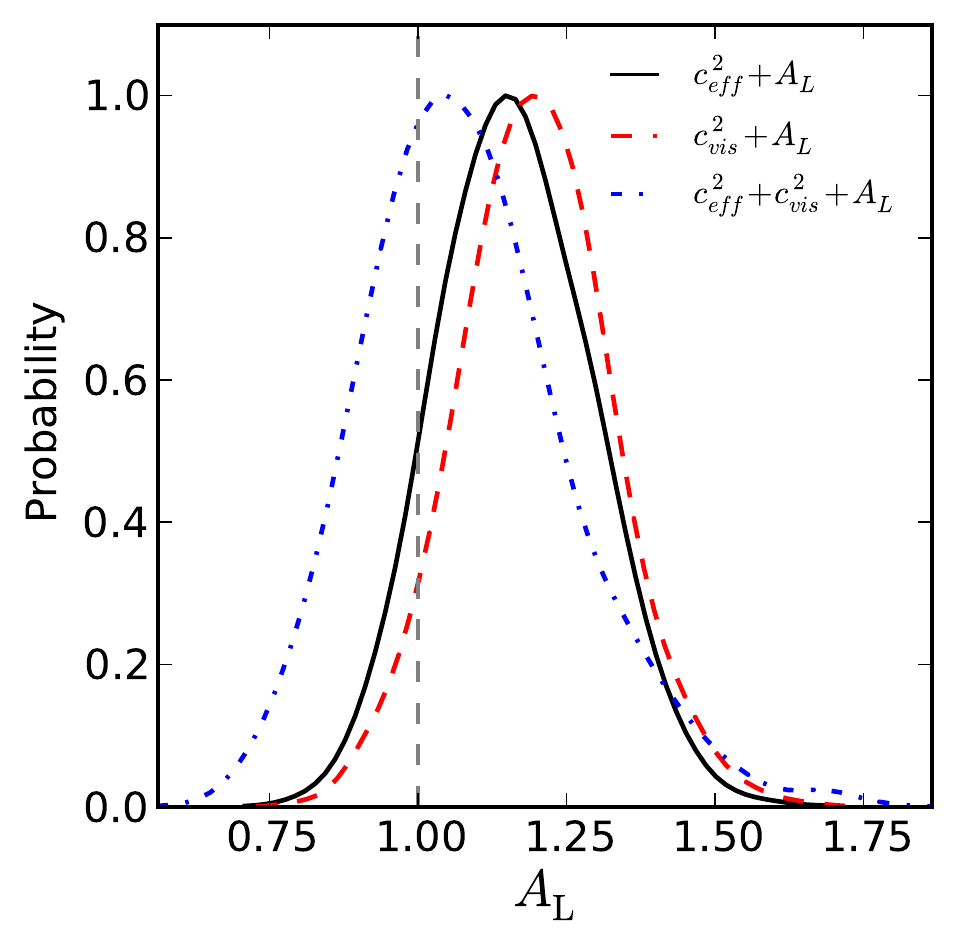}
\includegraphics[width=0.44\linewidth,keepaspectratio]{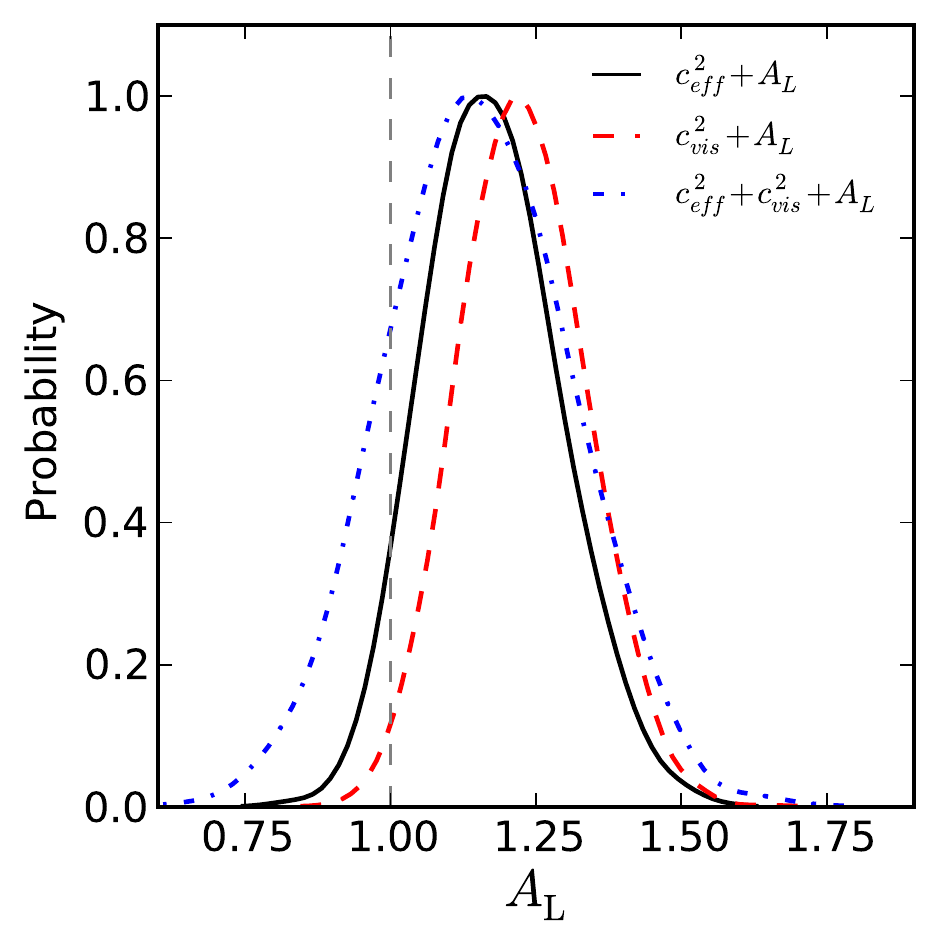}
\includegraphics[width=0.44\linewidth,keepaspectratio]{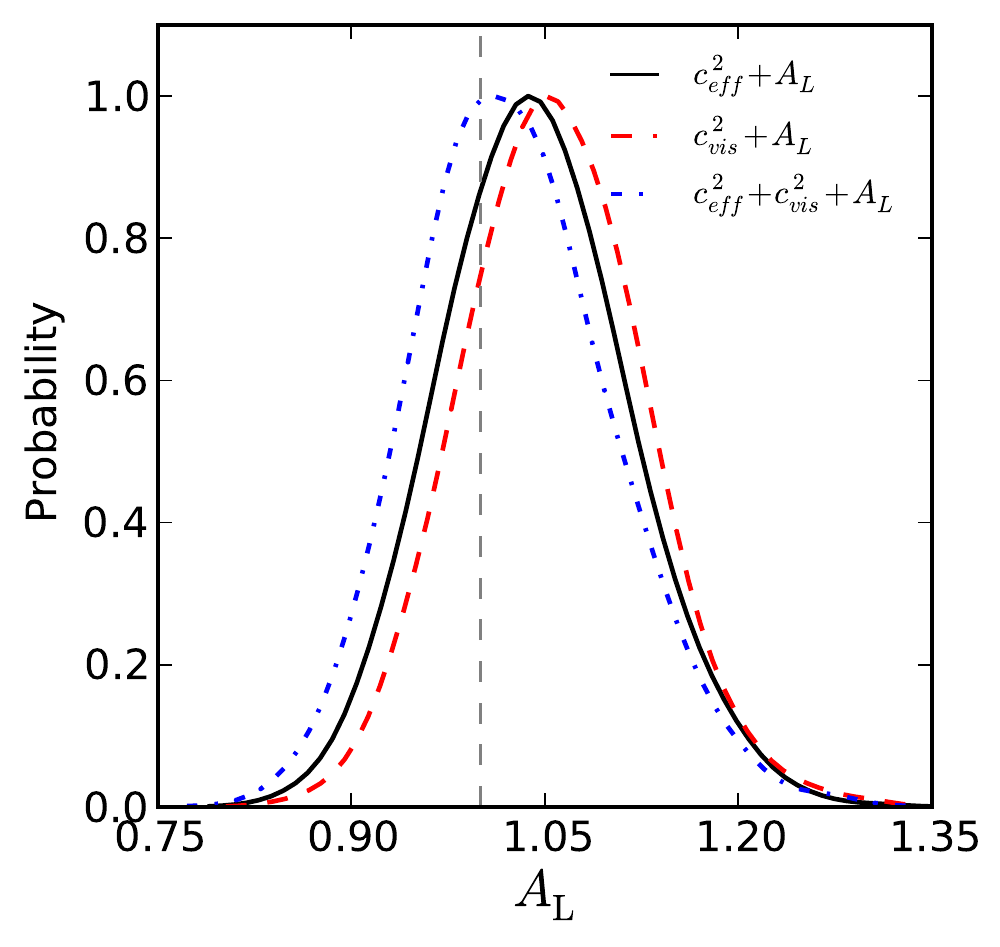} \\ 
\caption{One-dimensional posterior probabilities of the parameter $A_{\rm L}$ for the indicated models for $PlanckWP$ (upper left), $PlanckEX$ (upper right) and $PlanckL$ (bottom) datasets. The vertical dashed line indicates the expected value in the standard model. Note the different range in the $x$ axes for the $PlanckL$ dataset.}

\label{fig:Alens_1D}
\end{figure*}


\section{Results} \label{sec:results}
\subsection{Joint analysis}

We show our results in Tab.\ref{tab:paramsWP}, Tab.\ref{tab:paramsEX} and Tab.\ref{tab:paramsL} in the form of the $68 \%$ confidence level, {\textit i.e.,} the interval containing $68 \%$ 
of the total posterior probability centered on the mean. 

As we can see from Tab.\ref{tab:paramsWP}, Fig.\ref{fig:ceff_1D} and Fig.\ref{fig:cvis_1D} (upper left panels), allowing both $c_{\rm eff}^2$ and $c_{\rm vis}^2$ to vary produces posterior values in disagreement with the standard model. In particular, the constraints on $c_{\rm vis}^2$  point towards a value larger than $1/3$, being $c_{\rm vis}^2=0.60\pm0.18$ at $68 \%$ c.l., in tension with the standard value at about $1.5$ standard deviations. On the other hand, $c_{\rm eff}^2$ assumes in our analysis a value lower than the expected one: $c_{\rm eff}^2=0.304\pm0.013$
at $68 \%$ c.l., ruling out the standard value at more than $95 \%$ c.l.. 
The addition of BAO and HST datasets only results in an improvement of the constraining power (Tab.\ref{tab:paramsEX} and upper right panel in Fig.\ref{fig:ceff_1D} and Fig.\ref{fig:cvis_1D}). 

\begin{figure*}[htb!]
\begin{center}
\includegraphics[width=0.45\linewidth,keepaspectratio]{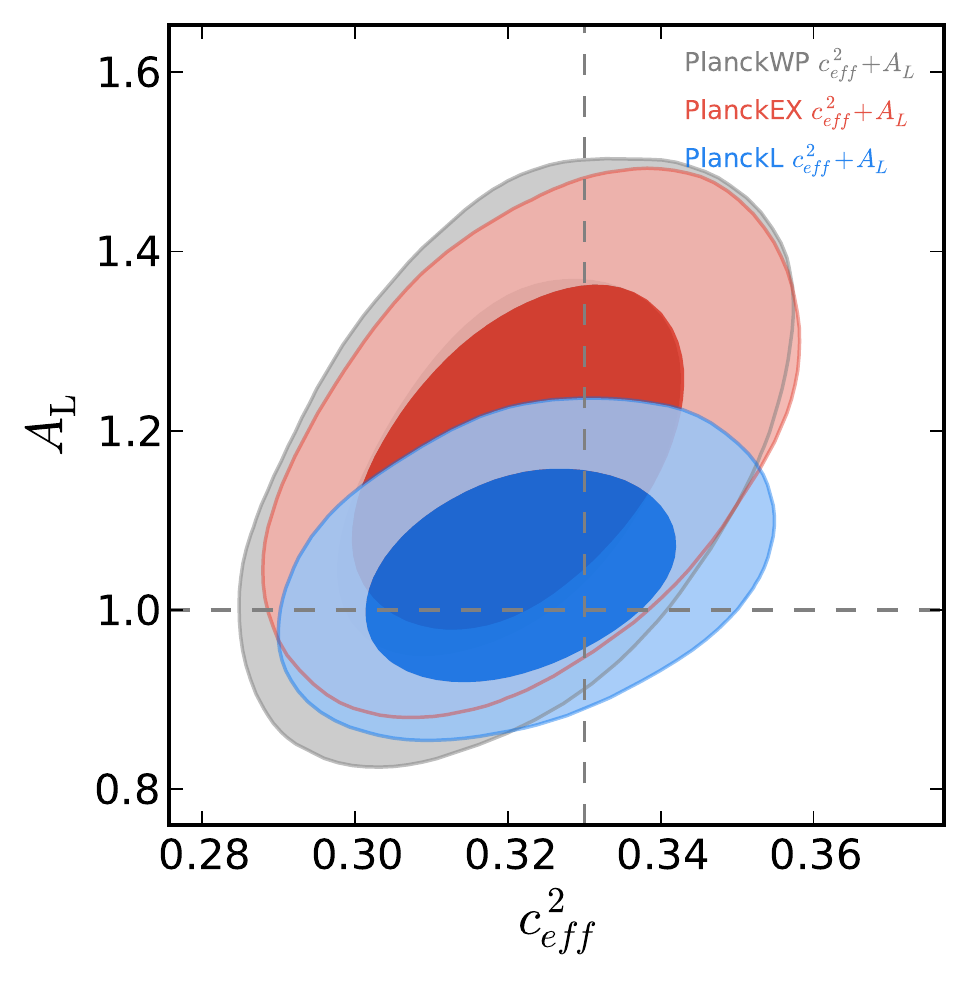}
\includegraphics[width=0.45\linewidth,keepaspectratio]{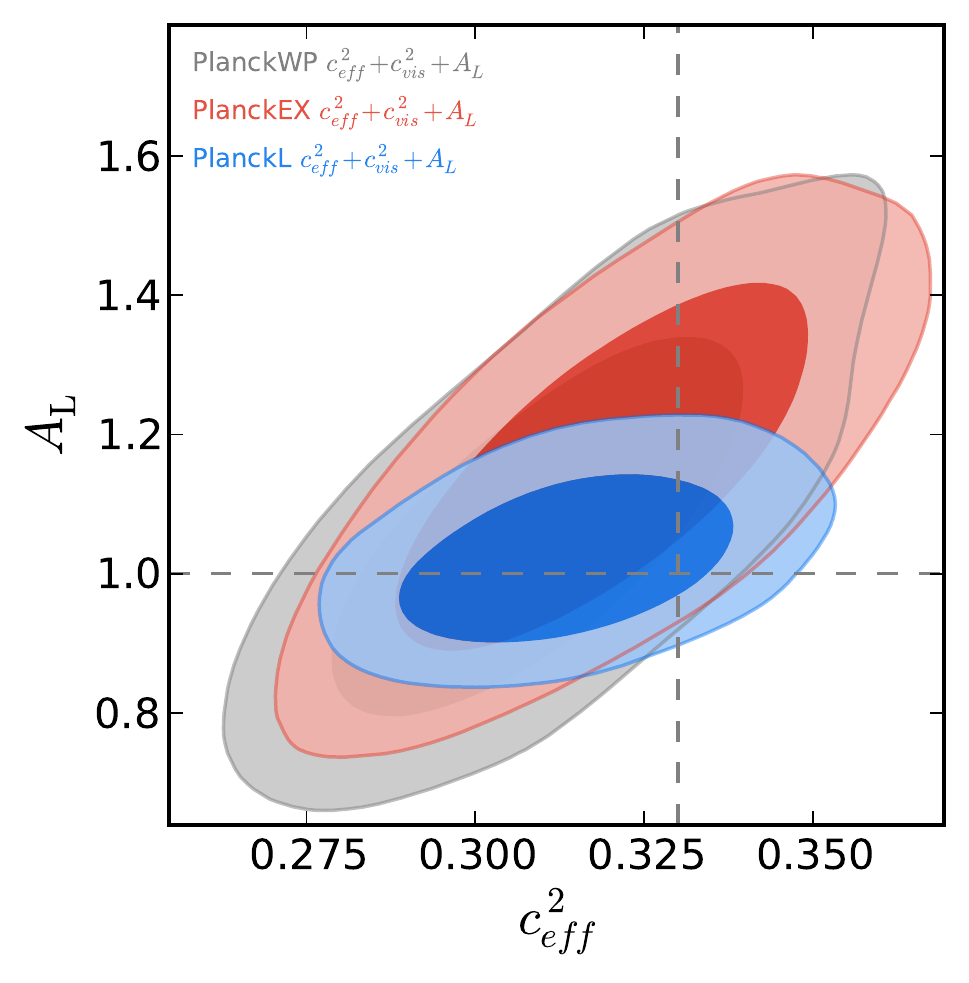} \\
\caption{Two-dimensional posterior probabilities in the $c_{\rm eff}^2-A_{\rm L}$ plane for the indicated datasets and models. The dashed lines indicate the expected values in the standard model.}

\label{fig:ceffAlens_2D}
\end{center}
\end{figure*}

\begin{figure*}[htb!]
\begin{center}
\includegraphics[width=0.45\linewidth,keepaspectratio]{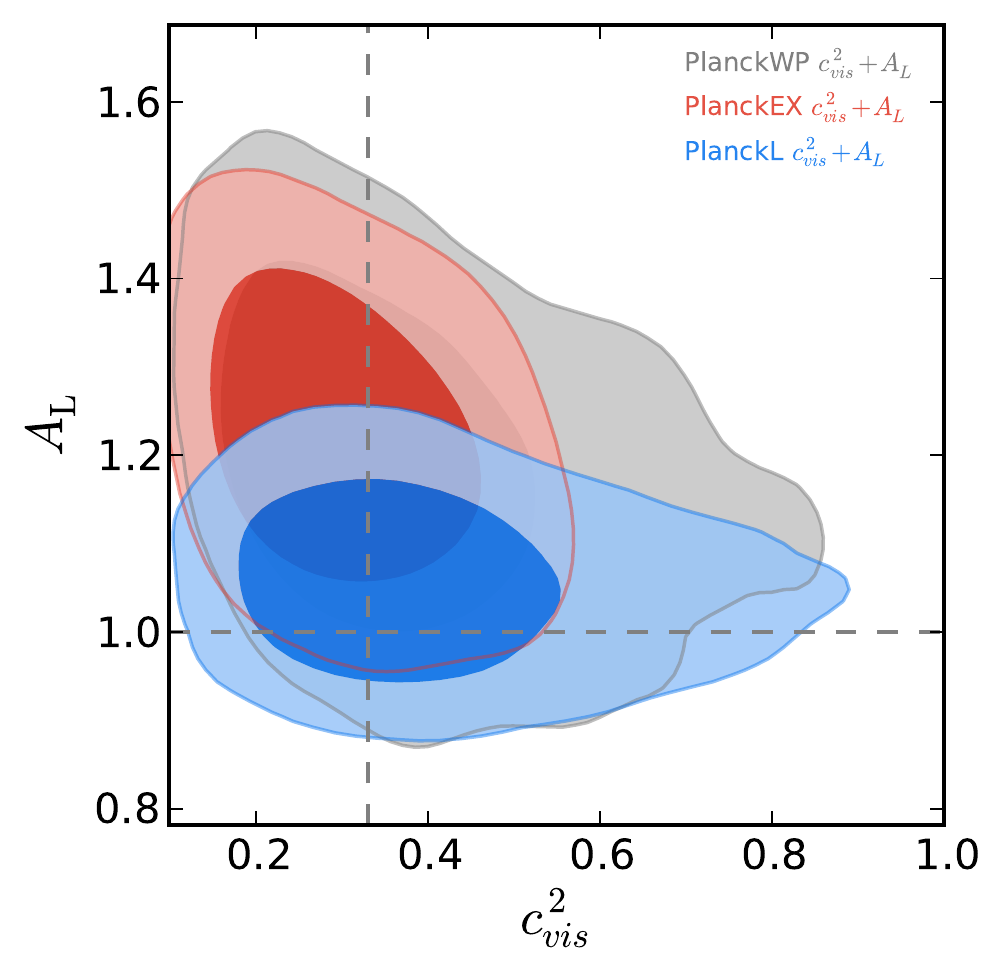}
\includegraphics[width=0.45\linewidth,keepaspectratio]{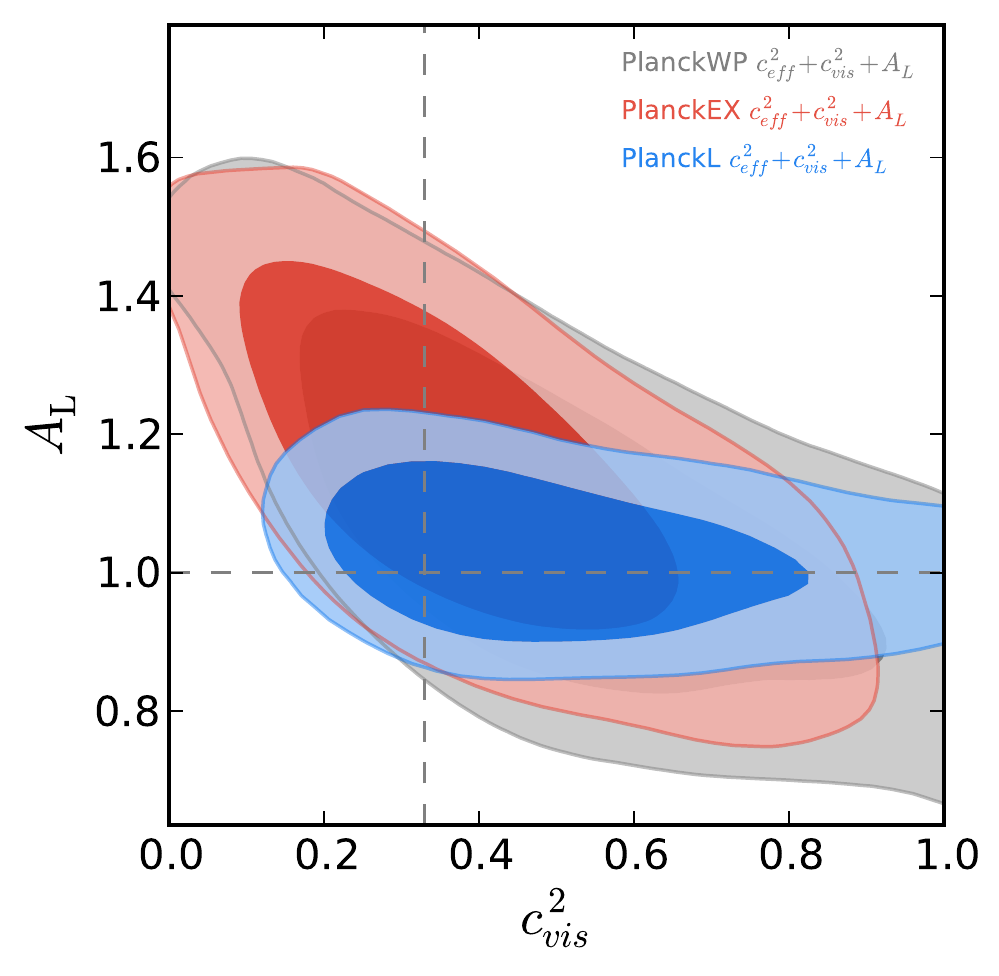} \\
\caption{Two-dimensional posterior probabilities in the $c_{\rm vis}^2-A_{\rm L}$ plane for the indicated datasets and models. The dashed lines indicate the expected values in the standard model.}

\label{fig:cvisAlens_2D}
\end{center}
\end{figure*}

The situation is significantly different when the lensing amplitude parameter is allowed to vary (see Tab.\ref{tab:paramsWP}, upper right panels in Fig.\ref{fig:ceff_1D}, Fig.\ref{fig:cvis_1D} and Fig.\ref{fig:Alens_1D}). In the $c_{\rm vis}^2+A_{\rm L}$ case, the standard value of the viscosity parameter is recovered ($c_{\rm vis}^2=0.35\pm0.12$); similarly, in the $c_{\rm eff}^2+A_{\rm L}$ case, $c_{\rm eff}^2$ is in agreement with the expected value ($c_{\rm eff}^2=0.321\pm0.014$). However, though the $A_{\rm L}=1$ case is now in more agreement
with the data, it is still in disagreement with the standard value 
at more than $1\sigma$ level (respectively, $A_{\rm L}=1.20\pm0.12$ and $A_{\rm L}=1.16\pm0.13$), showing a degeneracy between $A_{\rm L}$ and the clustering parameters $c_{\rm vis}^2$ and $c_{\rm eff}^2$ (see Fig.\ref{fig:ceffAlens_2D} and Fig.\ref{fig:cvisAlens_2D}). Also for these choices of parameters, the addition of BAO and HST allows to get tighter constraints as well (Tab.\ref{tab:paramsEX} and upper right panel in Fig.\ref{fig:ceff_1D} and Fig.\ref{fig:cvis_1D}).

Finally, when all the three parameters are allowed to vary, their posteriors are in good agreement (within $1\sigma$ c.l.) with the standard model: we find $c_{\rm eff}^2=0.311\pm0.019$, $c_{\rm vis}^2=0.51\pm0.22$ and $A_{\rm L}=1.08\pm0.18$. In this case, the addition of BAO and HST datasets slightly affects the peak location in the posterior distributions (see Tab.\ref{tab:paramsEX}, upper right panels in Fig.\ref{fig:ceff_1D}, Fig.\ref{fig:cvis_1D} and Fig.\ref{fig:Alens_1D}). In particular, it produces a $0.5 \, \sigma$ displacement of the $c_{\rm vis}^2$ mean towards a smaller (therefore more compatible with the expected) value. Conversely, it results in a $0.3\,\sigma$ shift of the $A_{\rm L}$ mean towards a higher value (less compatible with the expected, but more compatible with the $Planck$ collaboration finding). On the other hand, the magnitude of the displacement induced on the $c_{\rm eff}^2$ distribution (towards the expected value) is minimal ($<0.3\,\sigma$).

The results for the $PlanckL$ dataset are summarized in Tab.\ref{tab:paramsL}. The addition of the lensing power spectrum alleviates the disagreement respect to the standard values for the neutrino parameters (lower panels in Fig.\ref{fig:ceff_1D}, Fig.\ref{fig:cvis_1D} and Fig.\ref{fig:Alens_1D}). The greatest effect concerns the $\Lambda CDM + c_{\rm eff}^2 + c_{\rm vis}^2$ model, since we find $c_{\rm eff}^2 = 0.312\pm0.013$ and $c_{\rm vis}^2 = 0.52\pm0.18$ at 68\% c.l.. As we can see from Fig.\ref{fig:ceff_1D} and Fig.\ref{fig:cvis_1D}, allowing the lensing amplitude to vary produces a further shift towards the expected values in the standard scenario: $c_{\rm eff}^2 = c_{\rm vis}^2 = 1/3$ is compatible within 68\% c.l.. Finally, as far as the lensing amplitude is concerned, the addition of the lensing power spectrum results in a better agreement with the expected value, especially when all the three parameters vary jointly ($A_{\rm L} = 1.025\pm0.076$).

\subsection{Degeneracy}
\begin{figure*}[htb!]
\includegraphics[width=0.445\linewidth,keepaspectratio]{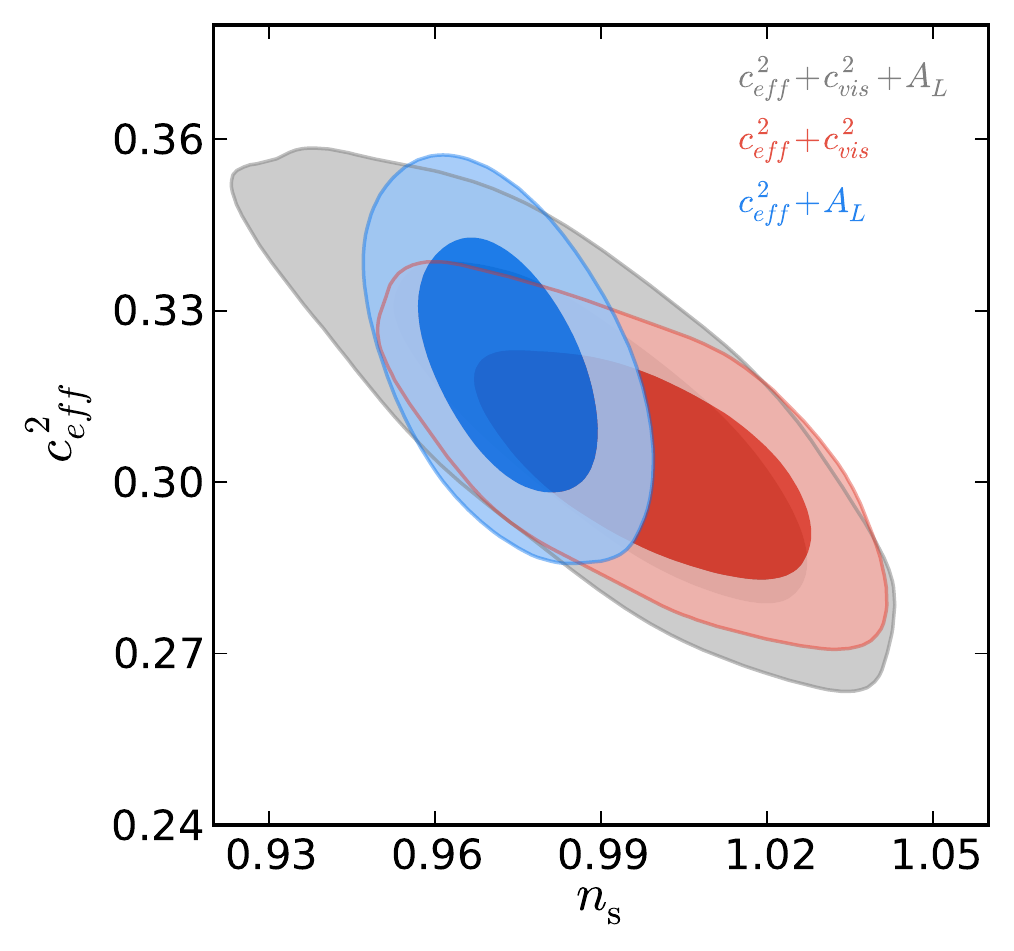}
\includegraphics[width=0.45\linewidth,keepaspectratio]{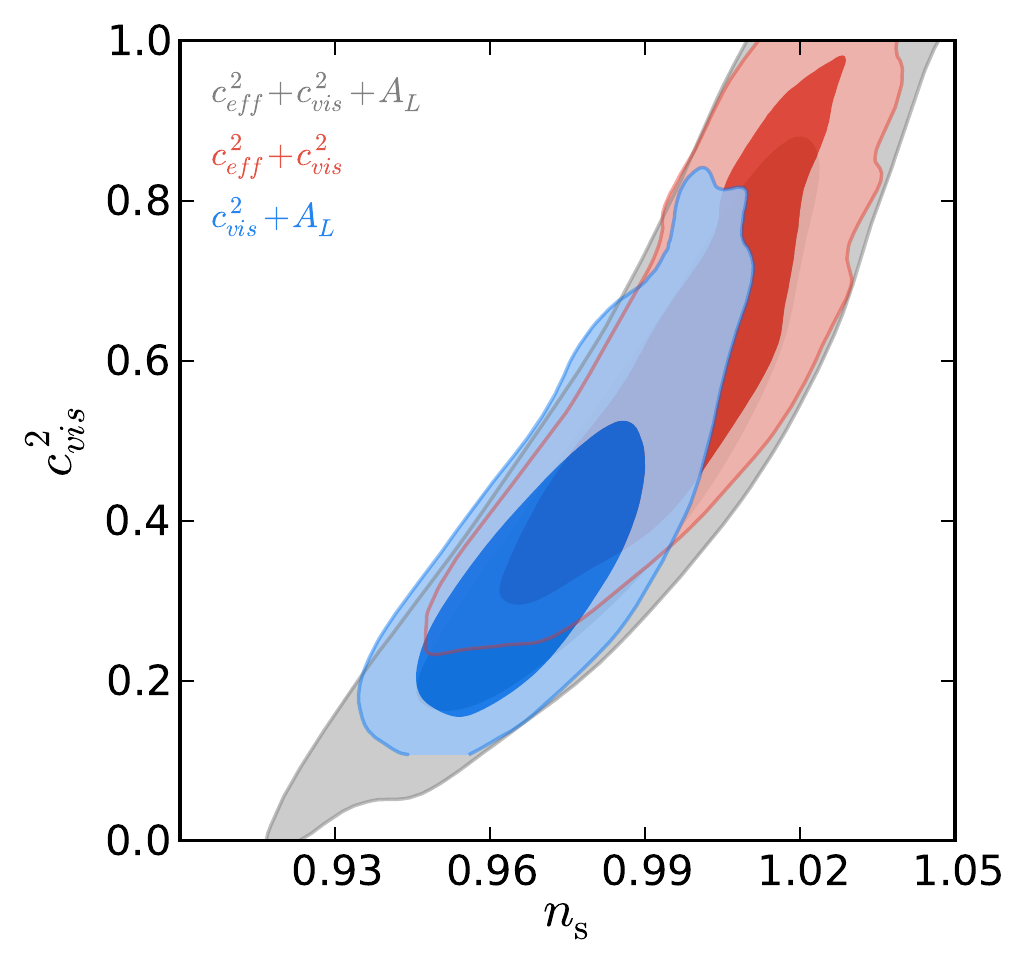} \\
\caption{Degeneracy between the clustering parameters $c_{\rm eff}^2$ (left) and $c_{\rm vis}^2$ and the scalar spectral index $n_S$ for the $PlanckWP$ dataset and the indicated models.}

\label{fig:ns}
\end{figure*}

\begin{figure*}[htb!]
\begin{center}
\includegraphics[width=0.45\linewidth,keepaspectratio]{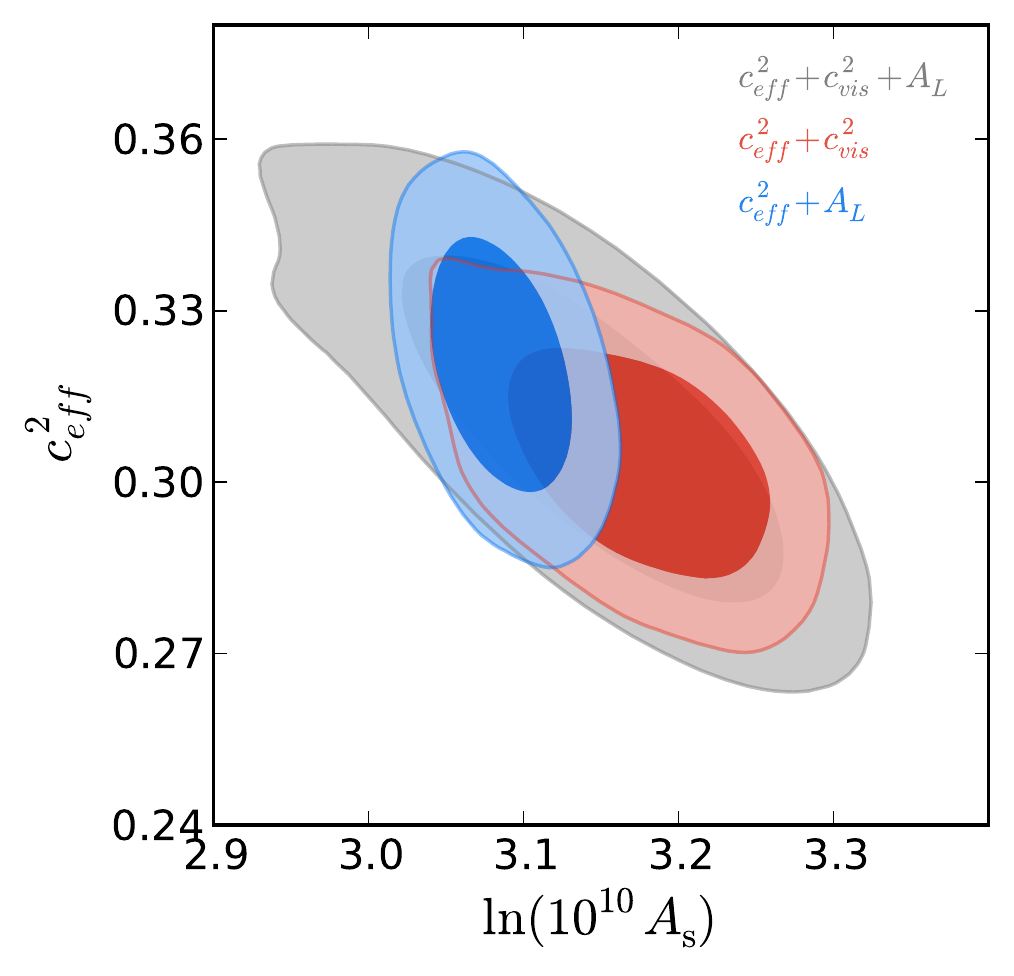}
\includegraphics[width=0.44\linewidth,keepaspectratio]{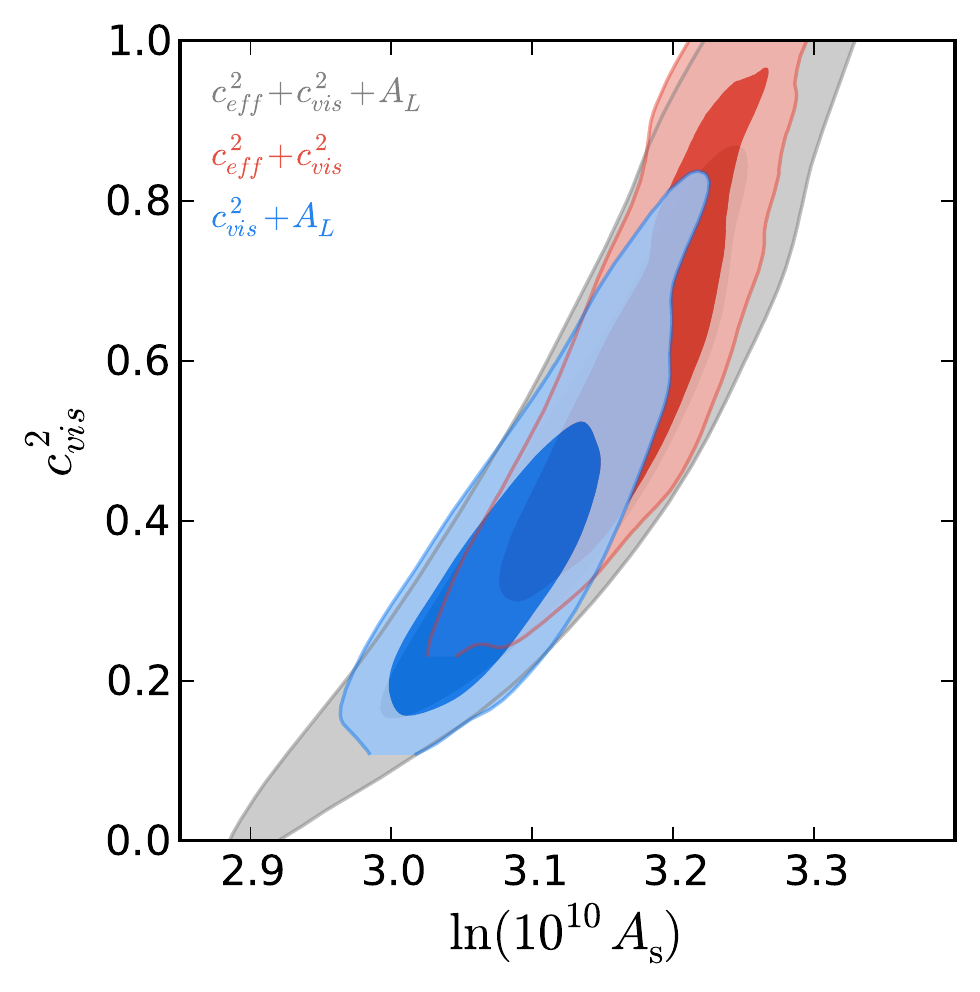} \\
\caption{Degeneracy between the clustering parameters $c_{\rm eff}^2$ (left) and $c_{\rm vis}^2$ and the scalar amplitude $A_S$ for the $PlanckWP$ dataset and the indicated models.}

\label{fig:logA}
\end{center}
\end{figure*}

As we can see from Tab.\ref{tab:paramsWP}, Tab.\ref{tab:paramsEX} and Tab.\ref{tab:paramsL}, varying the neutrino parameters also results in pronounced variations in other cosmologcal parameters, in particular the scalar spectral index $n_S$ and the scalar amplitude $A_S$. A similar analysis has been performed in \cite{Trotta:2004ty} considering just the ´´viscosity” $c_{\rm vis}^2$. Here, we would like to show the degeneracies between both the clustering parameters and the inflationary parameters, updated to more recent cosmological measurements. We report the two-dimensional posterior probabilities for the $PlanckWP$ dataset only. The conclusions are equivalent for the remaining two datasets.
As we can see from Fig.\ref{fig:ns} and Fig.\ref{fig:logA}, there is a negative (positive) correlation between $c_{\rm eff}^2$ ($c_{\rm vis}^2$) and the inflationary parameters. When the lensing amplitude varies together with one of the clustering parameters, the contours center at the standard value $c_{\rm eff}^2 = c_{\rm vis}^2 = 1/3$. However, while the choice of the model does not affect the direction of the degeneracy for $c_{\rm vis}^2$, the $\Lambda CDM +c_{\rm eff}^2 +A_{\rm L}$ model partly removes the degeneracy for $c_{\rm eff}^2$.

\begin{figure*}[htb!]
\begin{center}
\includegraphics[width=0.45\linewidth,keepaspectratio]{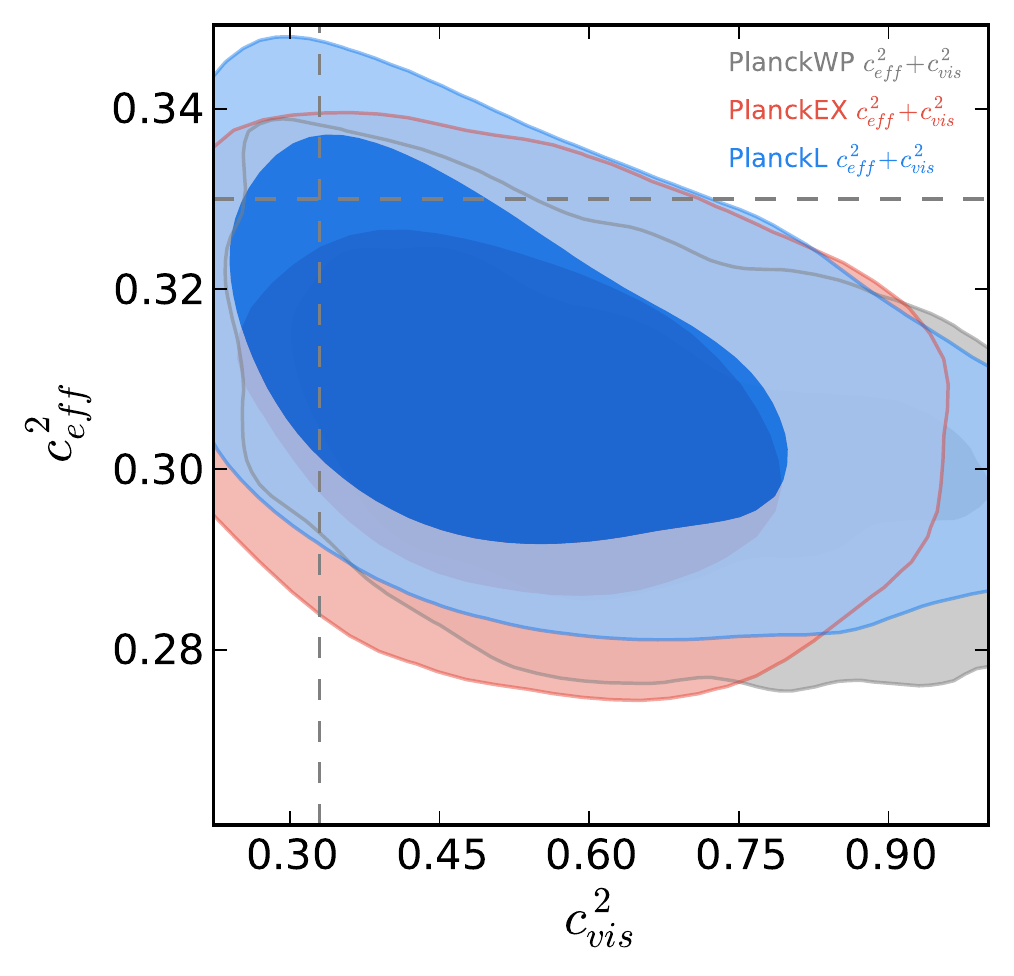}
\includegraphics[width=0.47\linewidth,keepaspectratio]{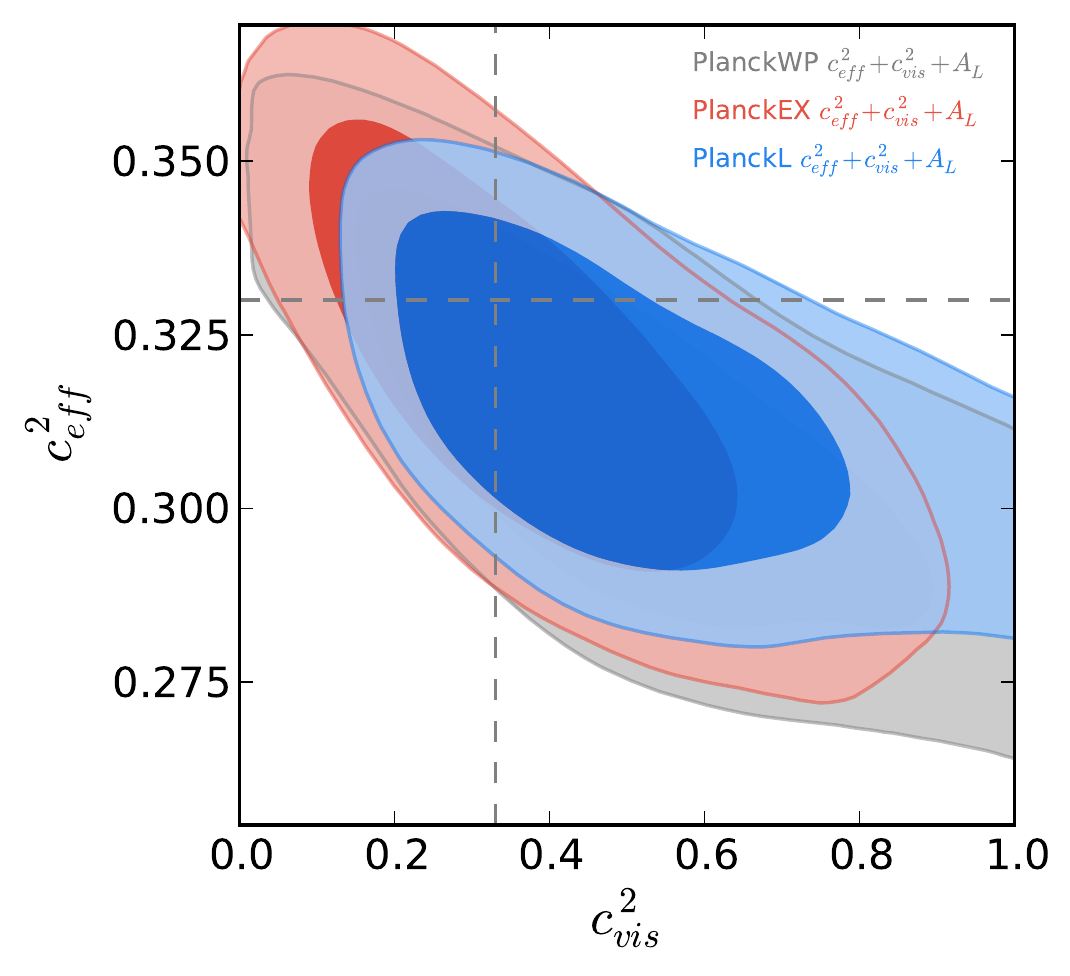} \\
\caption{Two-dimensional posterior probabilities in the $c_{\rm vis}^2-c_{\rm eff}^2$ plane for the indicated datasets and models. The dashed lines indicate the expected values in the standard model.}

\label{fig:ceffcvis_2D}
\end{center}
\end{figure*}

\section{Conclusions \label{sec:conclusions}}

In this paper we have presented new constraints on the clustering properties
of the neutrino background. We have found that the $Planck$ dataset hints
at anomalous values for these parameters with  $c_{\rm vis}^2=0.60\pm0.18$ at $68 \%$ c.l.
and $c_{\rm eff}^2=0.304\pm0.013$ at $68 \%$ c.l..
We have found a correlation between the neutrino parameters and the lensing amplitude of the 
temperature power spectrum $A_{\rm L}$.  When this parameter is allowed to vary we found a better
consistency with the standard model with $c_{\rm vis}^2=0.51\pm0.22$, 
$c_{\rm eff}^2=0.311\pm0.019$, and $A_{\rm L}=1.08\pm0.18$ at $68 \%$ c.l..
This result indicates that the anomalous large value of $A_{\rm L}$ measured by $Planck$ could be
connected to non-standard neutrino properties. 
Including additional datasets from Baryon Acoustic Oscillation surveys and the Hubble Space Telescope constraint on the
Hubble constant we obtain  $c_{\rm vis}^2=0.40\pm0.19$,  $c_{\rm eff}^2=0.319\pm0.019$, and $A_{\rm L}=1.15\pm0.17$ at $68 \%$ c.l..
The addition of the lensing power spectrum in the analysis allows to get a good agreement with the standard model as well: $c_{\rm vis}^2=0.50\pm0.19$, $c_{\rm eff}^2=0.314\pm0.015$, and $A_{\rm L}=1.025\pm0.076$ at $68 \%$ c.l..


\acknowledgments
It is a pleasure to thank Andrea Marchini, Olga Mena and 
Valentina Salvatelli for useful discussions.

\clearpage

\end{document}